\documentclass[runningheads,a4paper]{llncs}

\usepackage{amssymb}
\usepackage{graphicx}
\usepackage{color}
\usepackage{amsmath}
\usepackage{courier}

\usepackage{url}
\urldef{\mailsc}\path|dagmar.iber@bsse.ethz.ch|    
\newcommand{\keywords}[1]{\par\addvspace\baselineskip
\noindent\keywordname\enspace\ignorespaces#1}

\begin{document}

\mainmatter  

\title{Dynamic Image-Based Modelling of \\ Kidney Branching Morphogenesis}

\titlerunning{Image-Based Modelling of Branching Morphogenesis}

%
%
\author{Srivathsan Adivarahan$^1$
\and Denis Menshykau$^1$ 
\and  Odysse Michos$^2$
\and Dagmar Iber$^{1,3}$
\thanks{The authors acknowledge funding from the SNF Sinergia grant "Developmental engineering of endochondral ossification from mesenchymal stem cells", and a SystemsX RTD on Forebrain Development.}}
\authorrunning{Image-Based Modelling of Branching Morphogenesis}

\institute{
$^1$ Department for Biosystems Science and Engineering (D-BSSE), 
ETH Zurich, Basel, Switzerland \\
$^2$ Wellcome Trust Sanger Institute, Cambridge, UK \\
$^3$ Swiss Institute of Bioinformatics \\
\mailsc\\
\url{http://www.bsse.ethz.ch/cobi}}

%
%

\toctitle{Lecture Notes in Computer Science}
\tocauthor{Authors' Instructions}
\maketitle

\begin{abstract}
Kidney branching morphogenesis has been studied extensively, but the mechanism that defines the branch points is still elusive. Here we obtained a 2D movie of kidney branching morphogenesis in culture to test different models of branching morphogenesis with physiological growth dynamics. We carried out image segmentation and calculated the displacement fields between the frames. The models were subsequently solved on the 2D domain, that was extracted from the movie. We find that Turing patterns are sensitive to the initial conditions when solved on the epithelial shapes. A previously proposed diffusion-dependent geometry effect allowed us to reproduce the growth fields reasonably well, both for an inhibitor of branching that was produced in the epithelium, and for an inducer of branching that was produced in the mesenchyme. The latter could be represented by Glial-derived neurotrophic factor (GDNF), which is expressed in the mesenchyme and induces outgrowth of ureteric branches. Considering that the Turing model represents the interaction between the GDNF and its receptor RET very well and that the model reproduces the relevant expression patterns in developing wildtype and mutant kidneys, it is well possible that a combination of the Turing mechanism and the geometry effect control branching morphogenesis.

\end{abstract}

\section{Introduction}

Theoretical models have long been used to understand developmental patterning processes. Recent advances in imaging and computing allow us to develop increasingly realistic models of developmental pattern formation that can be validated with experimental data \cite{Iber:2012hm}. Such models open up new opportunities in that validated models can be used to clarify underlying mechanisms and to make predictions about further processes. The latter may enable a new field of \textit{in silico} genetics where mutations are tested computationally before creating a mouse mutant. The advantage of such an approach is that models may predict a lack of phenotype because of compensating regulatory interactions that would otherwise have been overlooked.  \textit{In silico} genetics can thus help to avoid inconclusive experiments.

\begin{figure}[b!]
\centering
\includegraphics[width=\textwidth]{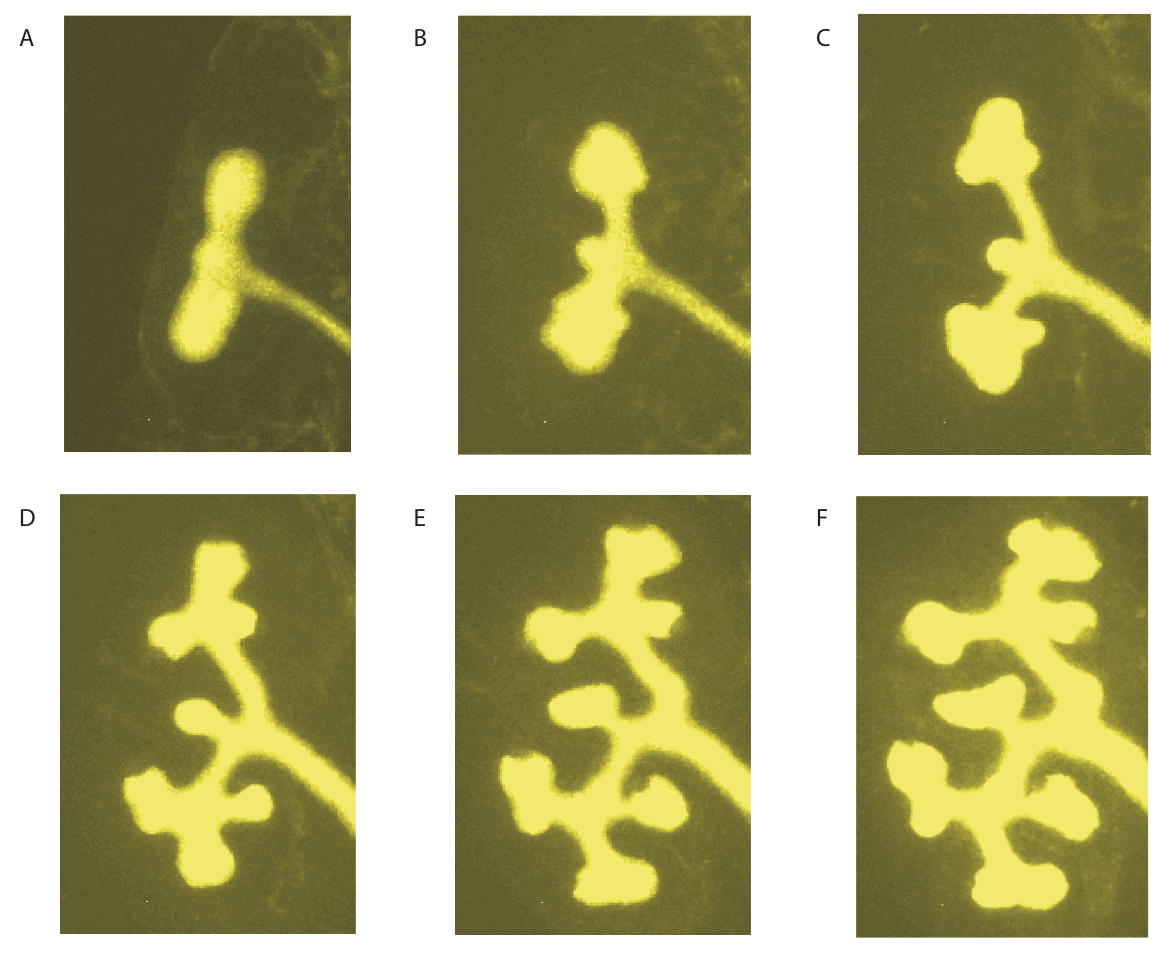}
\caption{Time course of kidney branching morphogenesis. The figure shows six out of 48 frames of a movie of kidney branching morphogenesis \emph{in vitro}. }
\label{fig:movie}
\end{figure}

Most of the information about developmental processes are image-based and patterns typically evolve on growing domains (Figure \ref{fig:movie}). The geometry of the domain, in turn, can greatly affect model predictions. It is therefore important to simulate models on such physiological, growing domains \cite{Iber:2013vf}. This requires the development and combination of suitable techniques. In this paper we describe the methodology to obtain the geometries and displacement fields of developing kidneys that are undergoing branching morphogenesis. These can then be used to test models that describe the processes that regulate branching by simulating the models on the extracted geometries and by comparing predicted signaling spots and embryonic growth field.

The kidney collecting ducts form via branching of an epithelial cell layer (Figure \ref{fig:movie}). During kidney development the ureteric bud invades the metanephric mesenchyme around embryonic day (E)10.5 \cite{Majumdar:2003tp}. It is currently not possible to image this branching process \textit{in utero}. We therefore obtained the data by culturing developing kidneys and by imaging the branching process over 48 hours. In culture, most branching events in the kidney are terminal bifurcations and to a lesser extent trifurcations, and only $6\%$ of all branching events are lateral branching events \cite{Watanabe:2004kr,Meyer:2004de,Costantini:2010p43730}. The branching pattern differs from the one observed in the embryo, which is likely the result of the different geometric constraints, but the core signaling mechanism should still be the same. The culture experiments should thus be adequately suited to test models for this core signaling mechanism.

\begin{figure}[t!]
\centering
\includegraphics[width=\textwidth]{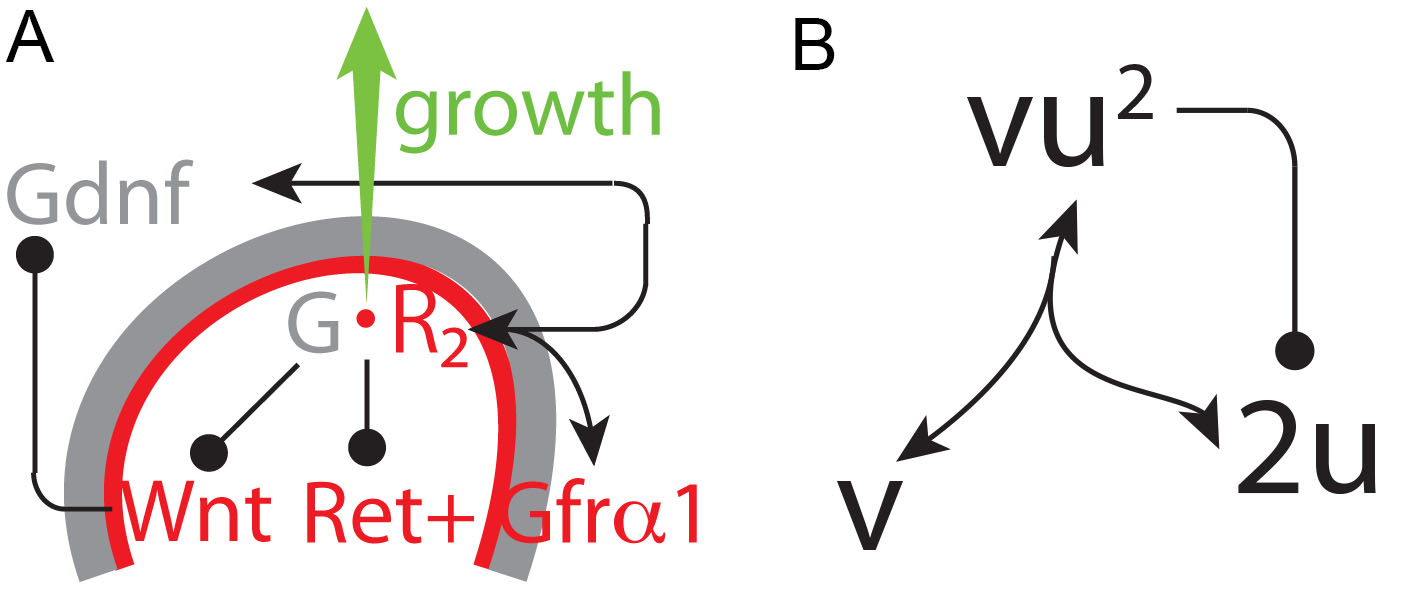}
\caption{The core Network regulating Kidney Branching Morphogenesis. (A) The dimer GDNF ($G$)  binds GFR$\alpha$1 and RET receptor ($R$) to form the  GDNF-receptor complex, $G\cdot R^2$. The complex induces the expression of the receptor, \textit{Ret}, and of \textit{Wnt11} ($W$). Moreover, signaling by the GDNF-RET receptor complex triggers bud outgrowth. Adapted from Figure 1A in \cite{Menshykau:nMxfL07C}. (B) Graphical representation of the ligand-receptor interactions in the simplified Schakenberg-type Turing model (Equations \ref{eq:Tur}). }
\label{fig:model}
\end{figure}

At the core of the mechanism controlling branching appears to be the TFG-beta family protein Glial-derived neurotrophic factor (GDNF) (Figure  \ref{fig:model}A). Thus beads soaked with GDNF induce the outgrowth of extra ureteric buds in kidney culture explants \cite{Costantini:2010p43730,Majumdar:2003tp,Treanor:1996dq,Pichel:1996en,Pepicelli:1997jz,Sanchez:1996cy}.  Based on the chemoattractive properties of  GDNF \cite{Tang:2002it,Tang:1998vf}, it was suggested that branching of the ureteric bud is caused by the attraction of the tips toward local sources of GDNF \cite{Sariola:2003jo}. Mice that do not express \textit{Gdnf}, or the GDNF receptor \textit{Ret}, or co-receptor GDNF family receptor alpha (\textit{Gfr$\alpha$1}), do not develop kidneys \cite{Costantini:2010p43730,Majumdar:2003tp,Treanor:1996dq,Pichel:1996en,Pepicelli:1997jz,Sanchez:1996cy,Schuchardt:1994hg}. GDNF signaling induces \textit{Wnt11} expression in the epithelial tip of the ureteric bud and WNT signaling up-regulates expression of \textit{Gdnf}  in the mesenchyme, which results in the establishment of an autoregulatory epithelial-mesenchymal feedback signaling loop. 

We have recently developed a model for the core network (Figure \ref{fig:model}A) of GDNF ($G$), RET ($R$), and WNT ($W$) \cite{Menshykau:nMxfL07C}, which in non-dimensional form reads
\begin{eqnarray}
\dot{G} &=&  \underbrace{\Delta  G}_{\text{diffusion}} +  \underbrace{\rho_\mathrm{G0} + \rho_\mathrm{G} \frac{W^2}{W^2 + 1}}_{\text{production}} - \underbrace{\delta_\mathrm{G} G}_{\text{degradation}} - \underbrace{\delta_\mathrm{C} R^2 G \nonumber}_{\text{complex formation}}\\
\dot{R} &=&  \underbrace{D_\mathrm{R} \Delta R}_{\text{diffusion}} +   \underbrace{\rho_\mathrm{R} + \nu R^2 G}_{\text{producation}} - \underbrace{\delta_\mathrm{R} R}_{\text{degradation}} - \underbrace{ 2\delta_\mathrm{C} R^2 G}_{\text{complex formation}}    \nonumber\\
\dot{W} &=&  \underbrace{D_\mathrm{W}  \Delta  W}_{\text{diffusion}} +    \underbrace{\rho_{W0} +\rho_\mathrm{W}   \frac{R^2 G}{R^2 G + 1}}_{\text{production}}-   \underbrace{\delta_\mathrm{W}W}_{\text{degradation}}.
\end{eqnarray}
When solved on a idealized 3D bud-shaped domain the model gives rise to GDNF-RET patterns that are reminiscent of, lateral branching events, bifurcations, and trifurcations. Much as reported for the embryo, the split concentration patterns as characteristic for bifurcations and trifurcations dominate in the model for physiological parameter values, while elongation and subsequent lateral branching are rather rare.  Further simulations on deforming domains showed that the split concentration profiles can support bifurcating and trifurcating outgrowth \cite{Menshykau:nMxfL07C}. \\

We previously noticed in a model for lung branching morphogenesis that the interaction between Sonic Hedgehog (SHH) and its receptor PTCH1 results in Schnakenberg-type reaction kinetics \cite{Menshykau:2012kg}. Similarly, we notice that the model for the biochemical interactions between GDNF and its receptors (Figure \ref{fig:model}A) reduces to Schnakenberg-type reaction kinetics of the form
\begin{eqnarray}\label{eq:Tur}
\frac{\partial u}{\partial \tau}&=&\Delta u +\gamma (a-u+u^2v) \nonumber \\
\frac{\partial v}{\partial \tau}&=&D\Delta v + \gamma (b-u^2v).
\end{eqnarray}
if we assume large concentrations of WNT, i.e. $W \gg 1$, a negligible receptor-independent decay rate, $\delta_G$, for GDNF, and $\nu \sim 3 \delta_C$. $u$ and $v$ then correspond to the receptor RET and its ligand GDNF respectively (Figure \ref{fig:model}B). Schnakenberg reaction kinetics \cite{Schnakenberg:1979td} can result in Turing pattern \cite{Gierer:1972vq}, i.e. in the emergence of stable pattern from noisy homogenous initial conditions, as a result of a diffusion-driven instability \cite{Turing:1952p868}. \\

Alternatively, it has been proposed that outgrowth of branches in the lung and mammary gland may be controlled by a diffusion-based geometry effect \cite{Gleghorn:2012el,Nelson:2006gn}. If ligand is produced only in part of a tissue, then diffusion will result in a higher concentration at the centre of the ligand-producing domain. If the ligand supports outgrowth then this could support budding. When analysed on epithelial shapes of developing chicken lungs, it was concluded that, due to the same diffusion effect, the lowest concentration was observed at the highly curved tips. The branching controlling factor would thus have to be an inhibitor of branching  \cite{Gleghorn:2012el}.\\

We tested both mechanisms by obtaining a 2D movie of cultured ureteric buds and by following their epithelial dynamics over time. We extracted the shapes and displacement fields and simulated our model on these physiological domains. We find that the Turing type pattern is unstable to the noise in the initial conditions when solved on the epithelial shapes. A Turing mechanism alone can thus not control branching morphogenesis in the kidney. The diffusion-based geometry effect allowed us to reproduce the measured growth fields reasonably well, as long as it was based on an inhibitor of branching, which was expressed in the epithelium,  or on an inducer of branching that was expressed in the mesenchyme. It is well possible that a combination of the Turing mechanism and the geometry effect control branching morphogenesis.

\section{Results}

To obtain the shapes of the ureteric bud during branching morphogenesis, E11.5 kidney rudiments were dissected and imaged as previously described \cite{Riccio:2012ew}. Kidneys were imaged every 60 minutes using the epifluorescence inverted microscope Nikon TE300. We obtained a total of 49 frames, six of which are shown in Figure \ref{fig:movie}. To solve our computational models on these dynamic geometries we  first segmented the images to obtain the boundary of the epithelium and calculated the displacement field between subsequent stages. The initial geometry and displacement fields were imported into the commercial FEM solver COMSOL Multiphysics to perform the simulations and parameter optimization.

\subsection{Image Segmentation and Border Extraction}

\begin{figure}[b!]
\centering
\includegraphics[width=\textwidth]{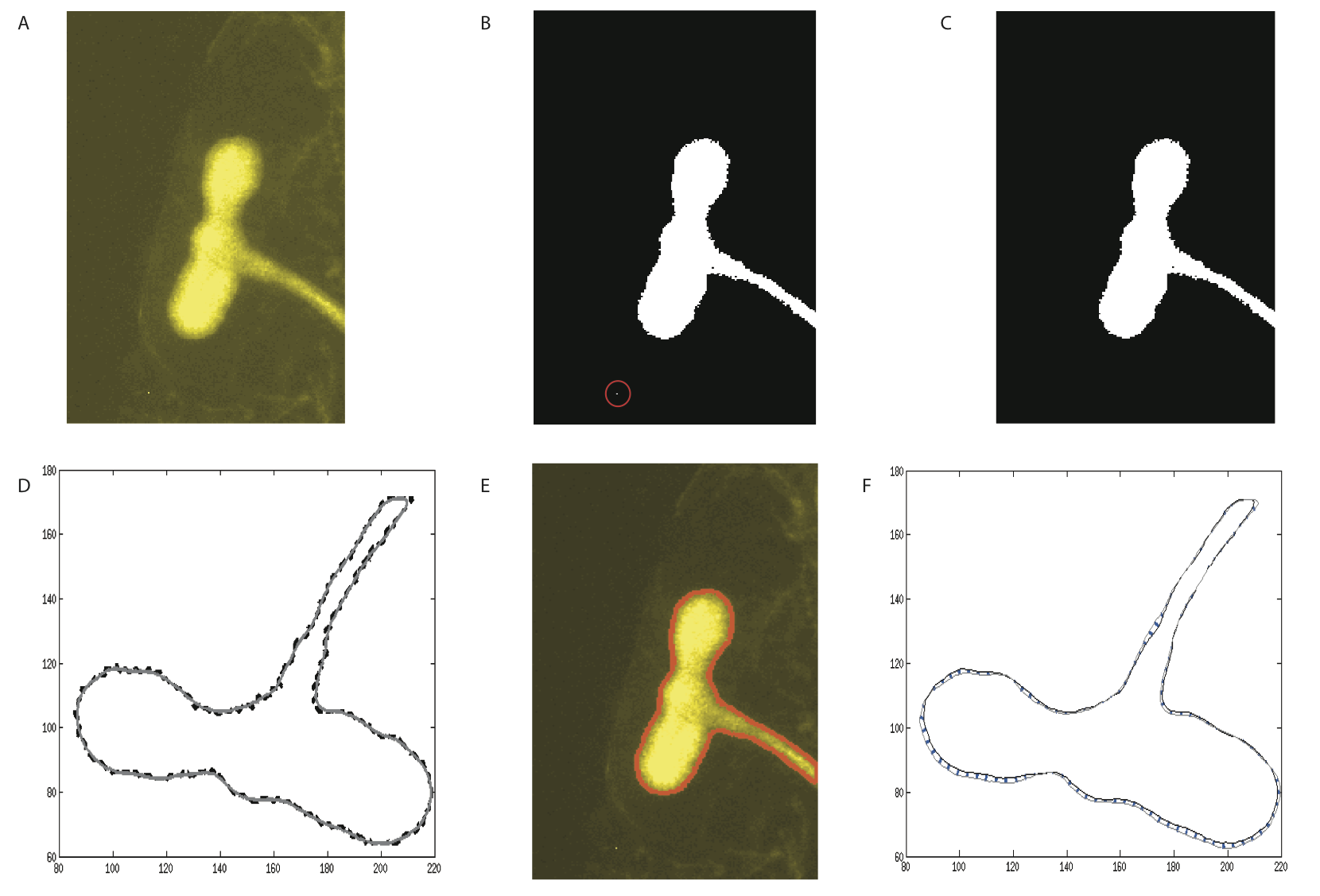}
\caption{Image Segmentation and Calculation of the Displacement Fields. 
(\textit{A}) An original image from the movie shown in Figure \ref{fig:movie}. (\textit{B}) Segmented image with a threshold filter.  Isolated points are still present; one such group of points has been marked by red circle. (C) Segmented image after the removal of isolated points. (\textit{D}) Extraction and smoothing of the boundary; the black dotted line shows the boundary obtained from C and the  grey line shows the smoothed boundary.   (\textit{E}) The smoothed boundary (in red) superimposed on the original image. (\textit{F}) Calculated displacement field between two images - the  blue lines represent the displacement vectors; the black and the grey lines represent the contours at two subsequent stages.}
\label{fig:segmentation}
\end{figure}

The images were segmented in MATLAB with a threshold based filter (Figure \ref{fig:segmentation}A,B). Prior to segmentation, the contrast of the image was increased with the built in MATLAB function \texttt{imadjust}. Next the images were segmented with a threshold filter. Threshold filters group pixels according to their intensity - pixels with intensities higher than a threshold value are assigned to the epithelium and those with intensity below a threshold value are assigned to the exterior. To apply threshold filters we used the MATLAB function \texttt{imb2bw}, which normalizes the intensity of each pixel prior to the application of the threshold filter. Threshold filters can wrongly assign islands of bright pixels to the kidney epithelium. To eliminate such small islands, we first labelled all separate objects with the MATLAB function \texttt{bwlabeln} and the object with the largest area was selected. (Figure \ref{fig:segmentation}C). We next extracted the border of the epithelium with the MATLAB function \texttt{bwboundaries} (Figure \ref{fig:segmentation}D). 

The extracted boundaries had to be smoothened before they could be used for simulations and further calculations. The smoothening was done using the MATLAB function \texttt{smooth} which uses a moving average method to smooth over the entire curve. Visual inspection confirmed that the extracted smoothened shape identifies the boundary of the kidney epithelium correctly (Figure \ref{fig:segmentation}E). The number of points in the extracted boundary was large, and were reduced to a set number using the interpolation function \texttt{interparc} \cite{D'Errico:2012:Online}.

\subsection{Calculation of a Displacement Field}

To simulate the signaling models on growing domains we needed to determine the displacement fields between the different stages. The displacement field between two consecutive stages was calculated by determining the minimum distance from each point on the curve at time $t$ to the curve at time $t+\Delta t$ using the MATLAB function \texttt{distance2curve} \cite{D'Errico:2012dist}. 


\subsection{Meshing and Simulations}

\begin{figure}[t!]
\centering
\includegraphics[width=\textwidth]{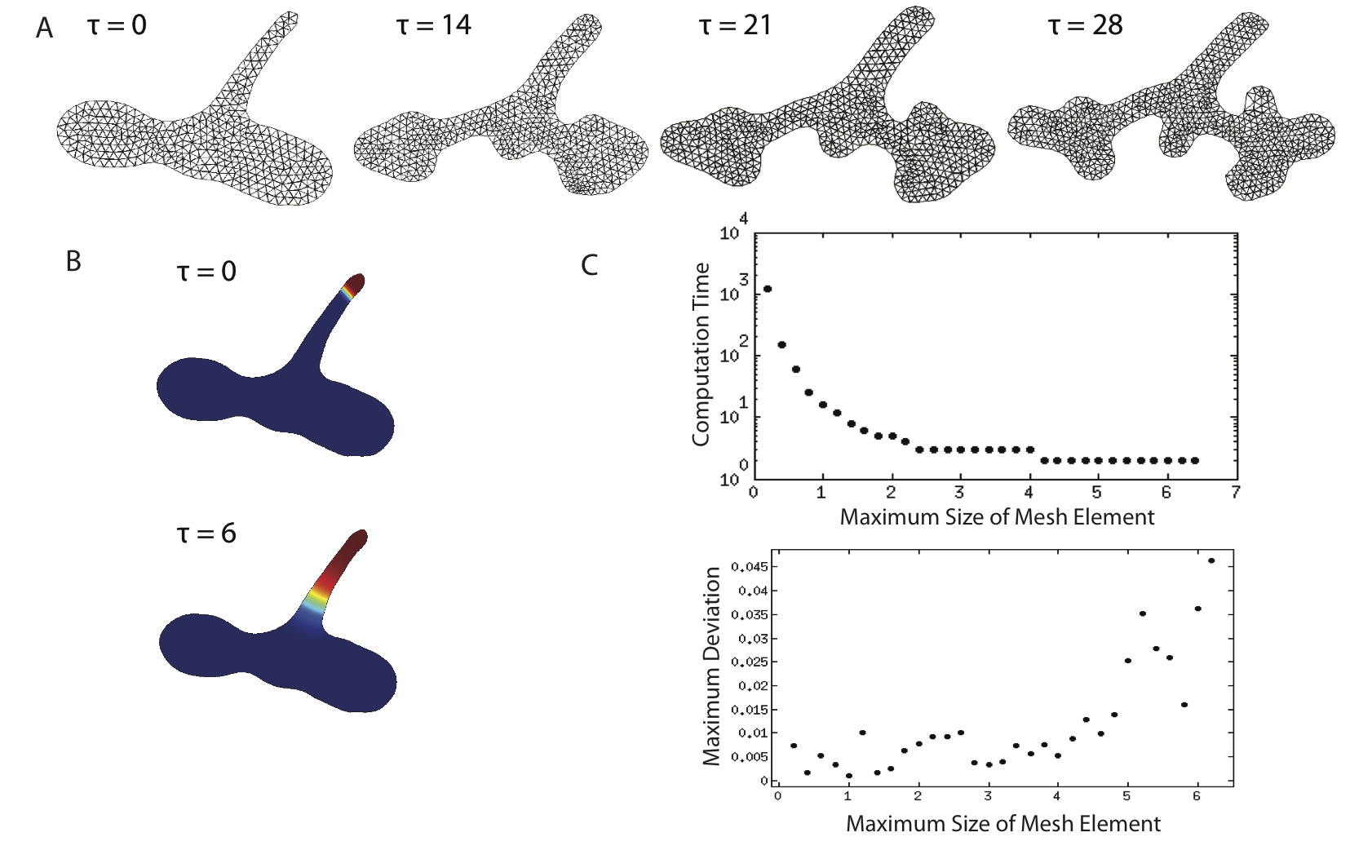}
\caption{Mesh Generation. (A) The meshed growing domain as generated in COMSOL Multiphysics at various time steps as indicated; the maximum size of the element was 3, minimum element quality 0.4 (B) Solution of a traveling wave equation (Equation \ref{eq:wave}) on a static domain at $\tau$=0 and $\tau$=6. (C) The upper plot shows the computational time for varying maximum size of mesh element (from 0.2 to 6.4) for $\tau$ = 6 while the lower plot shows the maximum deviation of the solution at $\tau$ = 6 for different mesh sizes. The deviation was calculated between a solution for a particular mesh size and the solution for a mesh with a mesh element size that was 0.2 greater than the current.} 
\label{fig:mesh}
\end{figure}

We next imported the curve that describes the initial shape of the epithelium  into the FEM-solver COMSOL Multiphysics, using the ASCII file format. COMSOL Multiphysics is a well-established software package and several studies confirm that COMSOL provides accurate solutions to reaction-diffusion equations both on constant\cite{cutress2010} and growing domains \cite{Carin2006,Thummler2007,Weddemann2008}. Details of how to efficiently implement these models in COMSOL have been described by us recently \cite{Menshykau:2012vg,Germann:bT_kMV7D}. The imported domain was meshed with a free triangular mesh (Figure \ref{fig:mesh}). The quality of the mesh can be assessed according to the following two parameters: mesh size and the ratio of the sides of the mesh elements. The linear size of the mesh should be much smaller than any feature of interest in the computational solution, i.e. if the gradient length scale in the model is 50 $\mu m$ then the linear size of the mesh should be at least several times less than 50 $\mu m$. Additionally, the ratio of the length of the shortest side to the longest side should be 0.1 or higher. 

Next the displacement field was imported into COMSOL and the domain was deformed accordingly. COMSOL Multiphysics uses the Arbitrary Lagrangian-Eulerian (ALE) formalism to solve PDEs on a deforming domain. Figure \ref{fig:mesh}A shows a sequence of meshes generated on a deforming domain. To confirm the convergence of the simulation, we solved a traveling wave equation of the form
\begin{equation}\label{eq:wave}
\frac{\partial u}{\partial \tau}=\Delta u + u(1-u)
\end{equation}
on a series of refined meshes (Figure \ref{fig:mesh}B). As the maximum mesh size decreased the maximum deviation in the solution decreased initially without greatly increasing the computation time (Figure \ref{fig:mesh}C).  As the mesh size was further decreased, the maximum deviation remained about constant while the computation time sharply increased. There is thus an optimal mesh size that needs to be defined for each particular model that is simulated on the domain.


\subsection{Kidney Branching Morphogenesis}

\begin{figure}
\centering
\includegraphics[width=\textwidth]{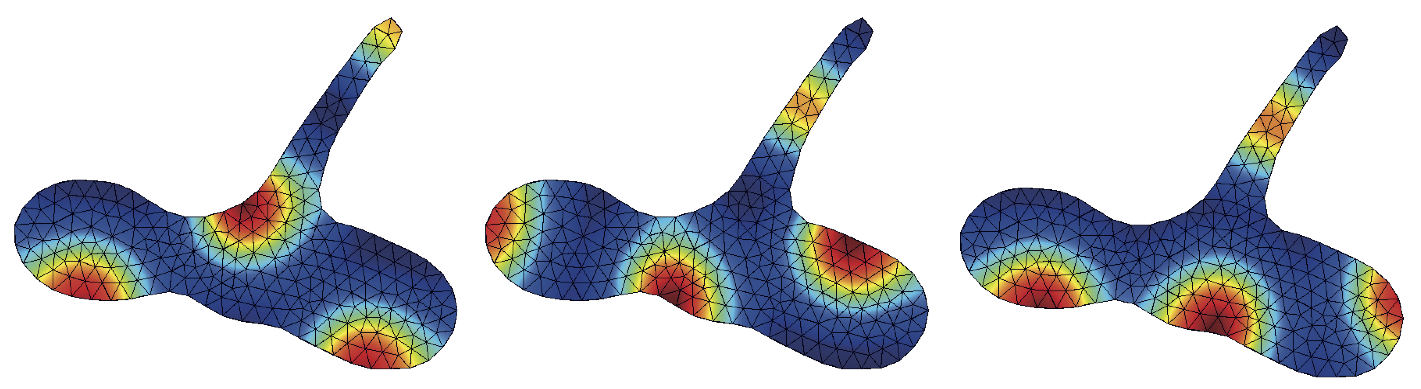}
\caption{Turing pattern on a static domain in the shape of embryonic kidney epithelium depends on noise in initial conditions. The three panels show the steady-state pattern of the receptor-ligand complex, $u^2v$ (rainbow color code: red - highest level, blue - lowest level). The three panels were computed with the same parameter set, but with different random initial conditions The parameter values used for generating the figure are: $a$ = 0.2, $b$ = 1.5 and $\gamma$ = 0.04, $D=100$. }
\label{fig:Turing}
\end{figure}

Depending on the choice of parameters Turing pattern can reproduce almost any pattern. We were interested whether we could find a parameter combination that would allow the model to reproduce the measured displacement field over time, while respecting all biological costraints. We started with a single frame of the extracted shape of the kidney epithelium from the movie (Figure \ref{fig:Turing}). When we simulated  the Schnakenberg Turing model (Equations \ref{eq:Tur}) on this shape we noticed that the emerging pattern depended on the initial conditions. Given the noise in these initial conditions many different patterns emerged for the same parameter set.  We therefore conclude that the Turing-based mechanism alone cannot explain the stereotyped, reliable pattern observed in the embryo. \\

A number of alternative mechanisms have previously been proposed to control branching morphogenesis in the lung. Most of these are based on the distance between the mesothelium and the epithelium, and thus cannot apply to the kidney. However, one mechanism relies on the particular tissue geometry  \cite{Gleghorn:2012el,Nelson:2006gn}, and we decided to test this one also for the kidney. 

\begin{figure}[t!]
\centering
\includegraphics[width=\textwidth]{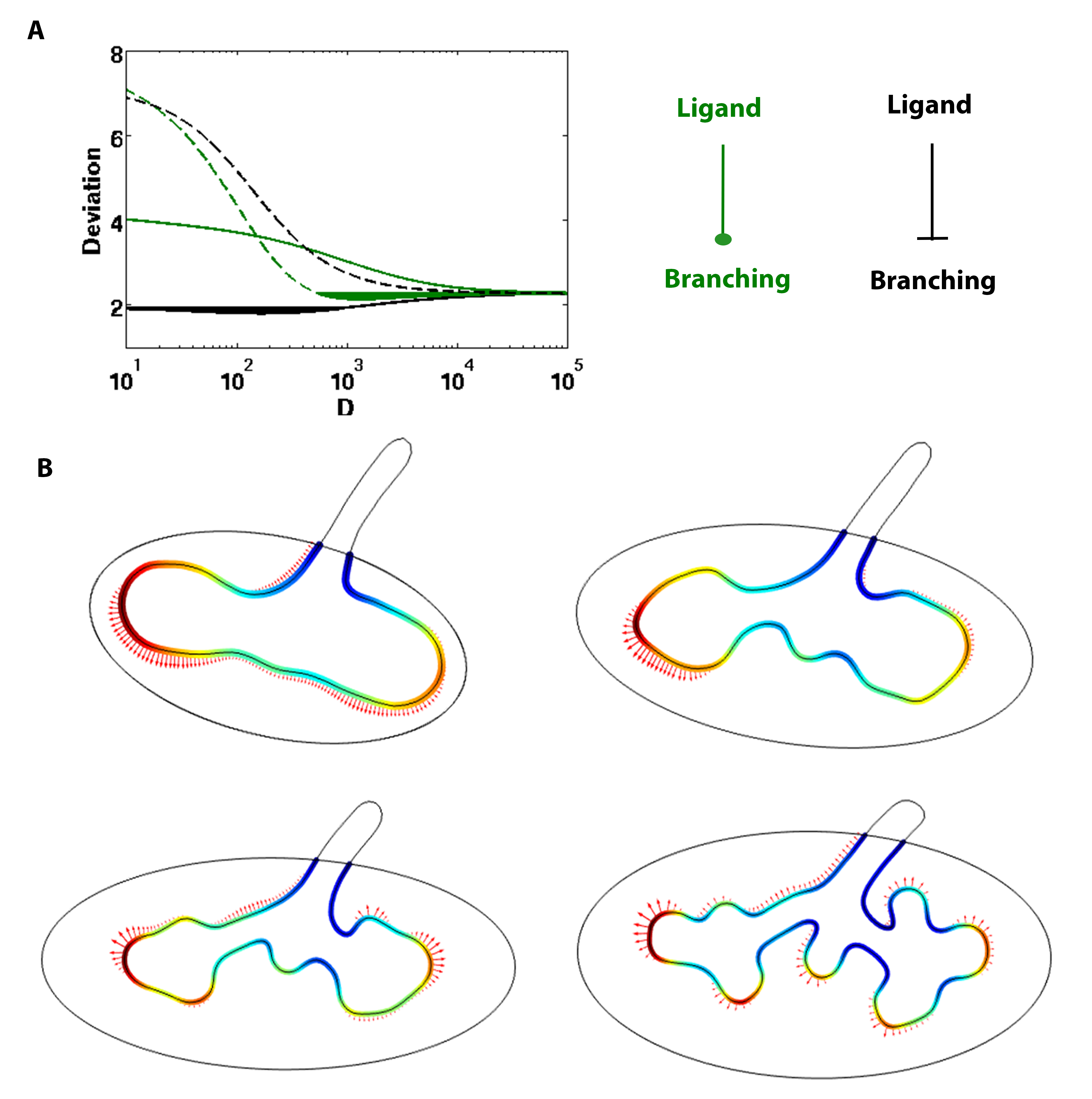}
\caption{A diffusion-based geometry effect results in patterns similar to those of the displacement field. (A) The deviation $\Delta$ (Eq. \ref{eq:deviate}) between signalling pattern and displacement field if $L$ is expressed either in the epithelium (solid lines) or in the mesenchyme (broken lines), and if $L$ acts either as activator (green lines) or inhibitor (black lines) of branch outgrowth. (B) The correspondence of the signalling effect and the displacement field for the best matching case, i.e. for $L$ expressed in the epithelium, acting as an inhibitor of branching, and $D=200$. The solid colours represent the value of $1/L$ (red - high to blue -low). The arrows mark the displacement field, with the length of the arrows indicating the strength of the displacement.}
\label{fig:Turing_mesenchyme}
\end{figure}

The mechanism requires that expression of the signalling factor is restricted to part of the tissue, and diffuses from there into the surrounding tissue. If ligand expression is restricted to the epithelium (and receptors to the mesenchyme) the model reads
\begin{eqnarray}\label{eq:L_Ep}
\textrm{Epithelium:} \quad & & \frac{\partial L}{\partial \tau} = D\Delta L + 1  \nonumber\\ 
\textrm{Mesenchyme:} \quad & &\frac{\partial L}{\partial \tau}=D\Delta L -L.
\end{eqnarray}
\textit{Gdnf} is expressed in the mesenchyme, and we therefore also wanted to study this case. The shape of the mesenchyme could not be extracted from the movies, and we therefore added an idealized domain in the shape of an ellipse to approximate the real shape of the mesenchyme. If ligand expression  is restricted to the mesenchyme (and receptors to the epithelium), then the model reads
\begin{eqnarray}\label{eq:L_Mes}
\textrm{Epithelium:} \quad & & \frac{\partial L}{\partial \tau} = D\Delta L -L  \nonumber\\ 
\textrm{Mesenchyme:} \quad & &\frac{\partial L}{\partial \tau}=D\Delta L +1.
\end{eqnarray}
Next we tested if the model could predict the areas of growth that were observed during kidney branching morphogenesis. To that end we adjusted the only parameter value in the model, $D$, to minimize the deviation, $\Delta$, between the computed signaling field and the registered displacement field based on the L2 distance (Euclidean distance), i.e. 
\begin{equation}\label{eq:deviate}
\Delta=\sqrt{\int\limits_L(|\overline{v}|-S)^2},
\end{equation}
$|\overline{v}|$ refers to normalized length of vectors of the displacement field, $S$ refers to the normalized computational signal. We used $S=L$ to model a ligand that induces branch outgrowth, and $S=1/L$ to model a ligand that inhibits branch outgrowth. The PDE models were solved on the kidney shapes of four separate stages (Figure \ref{fig:Turing_mesenchyme}B) for a wide range of the non-dimensional diffusion coefficient $D \in [10, \, 10^5]$. For each stage, 1000 parameter sets were sampled randomly from a logarithmic uniform distribution within these ranges.  \\

The lowest deviation was obtained for the model where the ligand $L$ was expressed in the epithelium and acted as an inhibitor of branch outgrowth (Figure  \ref{fig:Turing_mesenchyme}A, black, solid line). The best fitting pattern matches the observed growth field quite well, though not perfectly (Figure \ref{fig:Turing_mesenchyme}B). The second closest match was obtained when an activator of branching was expressed in the mesenchyme, which could be represented by GDNF. The other two cases did not provide a good match. The observed displacement of the stalk could not be captured by the model. However, we note that according to experimental observations (at least a later stages) the receptor \textit{Ret} is not expressed in the stalk \cite{Pachnis:1993um}; this displacement must thus be the result of other processes than GDNF/RET signaling. 

\begin{figure}[t!]
\centering
\includegraphics[width=\textwidth]{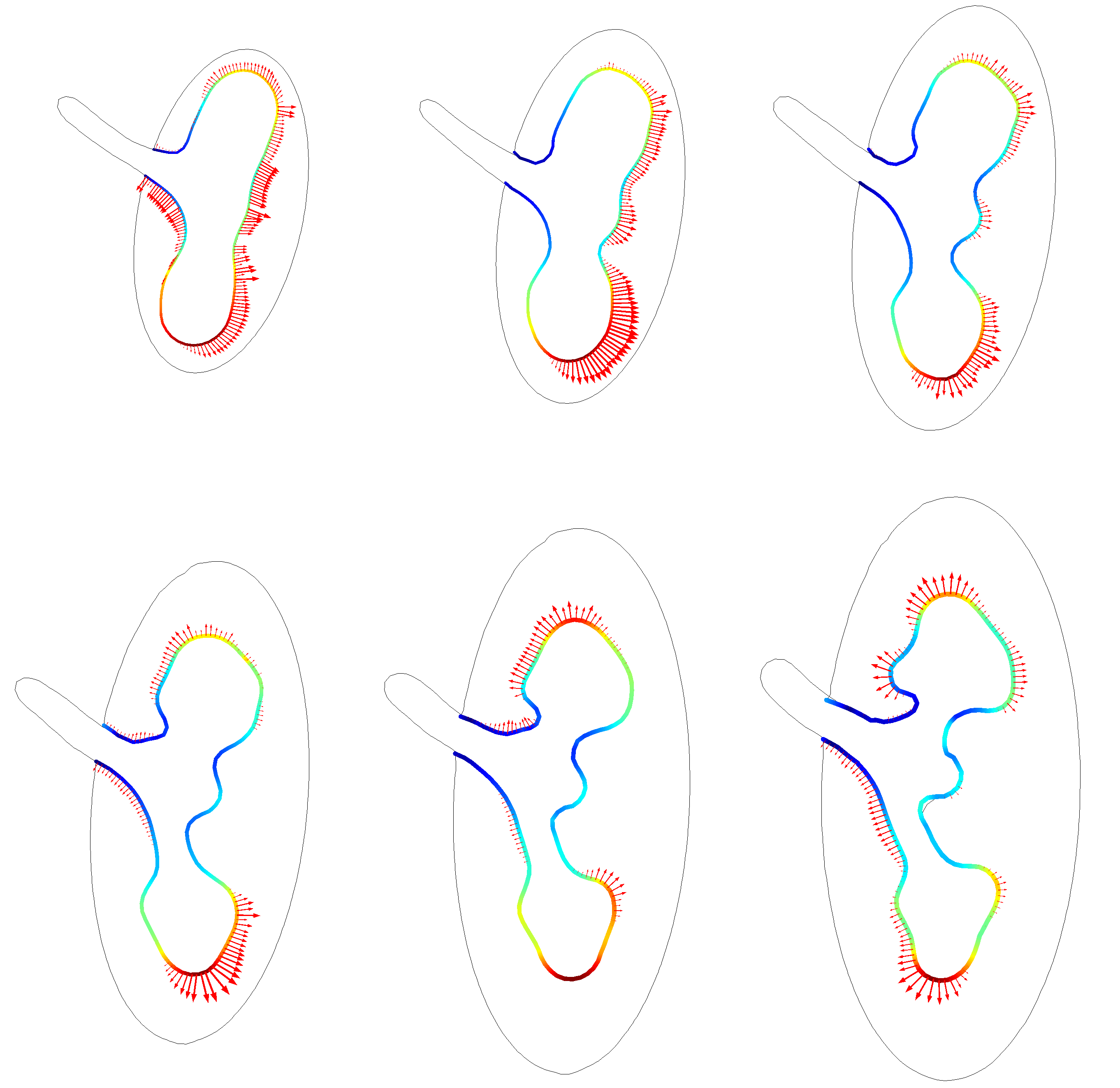}
\caption{Simulations on a continuously deforming domain, according to the measured displacement field. The correspondence of the signalling effect and the displacement field for the case of $L$ expressed in the epithelium, acting as an inhibitor of branching, and $D=200$. The solid colours represent the value of $1/L$ (red - high to blue -low). The arrows mark the displacement field, with the length of the arrows indicating the strength of the displacement.}
\label{fig:Turing_dynam}
\end{figure}

 Due to high computational cost we were unable to perform the optimization on the deforming domain; thus all results discussed so far were obtained on a series of static frames. To test whether the resulting pattern would be stable on a growing, deforming domain, we ran a simulation with the best fitting parameter value ($D=200$) on the recorded kidney movie, i.e. we started with the first frame and deformed the domain according to the measured displacement field of the kidney explants (Figure \ref{fig:Turing_dynam}). The distribution of the signalling activity that we obtained on a series of static frames (Figure  \ref{fig:Turing_mesenchyme}B) is indeed similar to the one obtained on a growing domain (Figure \ref{fig:Turing_dynam}).

\section{Discussion}

Mathematical models can help with the understanding of biological complexity if thoroughly rooted in experimental data. Most data in developmental biology is image based. In this contribution we used 2D movies that document branching of cultured kidneys to test a mathematical model for branching morphogenesis. We used MATLAB-based functions to extract the shapes and to calculate the displacement fields between frames. The shapes and displacement fields were subsequently imported into COMSOL to simulate the model on physiological geometries. Our previously proposed model for the core signaling mechanism is based on Schnakenberg reaction kinetics that can give rise to Turing pattern \cite{Menshykau:nMxfL07C}. When simulating the model on the extracted epithelial shapes we noticed that the pattern were sensitive to the noisy initial conditions, which rules out such mechanism for robust pattern formation.

We next tested a diffusion-based geometry effect \cite{Gleghorn:2012el,Nelson:2006gn}, which reproduced the growth fields of the cultured kidneys reasonably well. Much as in previous studies in the lung and mammary gland \cite{Gleghorn:2012el,Nelson:2006gn}, when the ligand was produced only in the epithelium, it needed to be an inhibitor of branching to explain the observed growth fields. On the other hand, if produced in the mesenchyme, it had to be an inducer of outgrowth. The latter could be represented by GDNF, an inducer of the outgrowth of the ureteric bud. 

While the Turing mechanism was unstable when solved on the epithelial domain, we note that it is well conceivable that both the Turing mechanism and the geometry effect act together during branching morphogenesis. After all, the Turing mechanism allowed us to reproduce the phenotype of all relevant, published mutants when solved on idealized domains \cite{Menshykau:nMxfL07C}. Further analysis of movies that include both the epithelium and the mesenchyme will be important to address this. 

The here-described methods permit the analysis of 2D movies. To further enhance the power of the analysis it would be valuable to obtain image frames in 3D rather than 2D. The calculation of the displacement field is more difficult in 3D. Several softwares are available to support morphing between 3D structures.  The software package AMIRA employs the landmark-based Bookstein algorithm \cite{Bookstein:1989wl}, which uses paired thin-plate splines to interpolate surfaces over landmarks defined on a pair of surfaces. The landmark points need to be placed by hand on the 3D geometries to identify corresponding points on the pair of surfaces. This is both time consuming and limits the accuracy of the reconstructed 4D series, in particular if the geometries are complex as is the case during kidney branching morphogenesis and if single frames are further apart. Further developments are clearly needed to enable a faster and more accurate reconstruction of 4D datasets. Similarly, computational methods need to be improved to facilitate the solution of computational models on complex, growing domains, that comprise several subdomains (tissue layers). While this is feasible, it is currently computationally very expensive, which makes it difficult to screen larger parameter spaces.

\bibliographystyle{plos2009}


\begin{thebibliography}{10}
\providecommand{\url}[1]{\texttt{#1}}
\providecommand{\urlprefix}{URL }
\expandafter\ifx\csname urlstyle\endcsname\relax
  \providecommand{\doi}[1]{doi:\discretionary{}{}{}#1}\else
  \providecommand{\doi}{doi:\discretionary{}{}{}\begingroup
  \urlstyle{rm}\Url}\fi
\providecommand{\bibAnnoteFile}[1]{%
  \IfFileExists{#1}{\begin{quotation}\noindent\textsc{Key:} #1\\
  \textsc{Annotation:}\ \input{#1}\end{quotation}}{}}
\providecommand{\bibAnnote}[2]{%
  \begin{quotation}\noindent\textsc{Key:} #1\\
  \textsc{Annotation:}\ #2\end{quotation}}
\providecommand{\eprint}[2][]{\url{#2}}

\bibitem{Iber:2012hm}
Iber D, Zeller R (2012) {Making sense-data-based simulations of vertebrate limb
  development.}
\newblock Curr Opin Genet Dev 22: 570--577.
\input{Iber:2012hm}

\bibitem{Iber:2013vf}
Iber D, Tanaka S, Fried P, Germann P, Menshykau D (2013) {Simulating Tissue
  Morphogenesis and Signaling }.
\newblock In: Nelson CM, editor, Tissue Morphogenesis: Methods and Protocols,
  Methods in Molecular Biology (Springer).
\input{Iber:2013vf}

\bibitem{Majumdar:2003tp}
Majumdar A, Vainio S, Kispert A, McMahon J, McMahon AP (2003) {Wnt11 and
  Ret/Gdnf pathways cooperate in regulating ureteric branching during
  metanephric kidney development.}
\newblock Development (Cambridge, England) 130: 3175--3185.
\input{Majumdar:2003tp}

\bibitem{Watanabe:2004kr}
Watanabe T, Costantini F (2004) {Real-time analysis of ureteric bud branching
  morphogenesis in vitro.}
\newblock Developmental Biology 271: 98--108.
\input{Watanabe:2004kr}

\bibitem{Meyer:2004de}
Meyer TN, Schwesinger C, Bush KT, Stuart RO, Rose DW, et~al. (2004)
  {Spatiotemporal regulation of morphogenetic molecules during in vitro
  branching of the isolated ureteric bud: toward a model of branching through
  budding in the developing kidney.}
\newblock Developmental Biology 275: 44--67.
\input{Meyer:2004de}

\bibitem{Costantini:2010p43730}
Costantini F, Kopan R (2010) {Patterning a complex organ: branching
  morphogenesis and nephron segmentation in kidney development.}
\newblock Dev Cell 18: 698--712.
\input{Costantini:2010p43730}

\bibitem{Treanor:1996dq}
Treanor JJ, Goodman L, de~Sauvage F, Stone DM, Poulsen KT, et~al. (1996)
  {Characterization of a multicomponent receptor for GDNF.}
\newblock Nature 382: 80--83.
\input{Treanor:1996dq}

\bibitem{Pichel:1996en}
Pichel JG, Shen L, Sheng HZ, Granholm AC, Drago J, et~al. (1996) {Defects in
  enteric innervation and kidney development in mice lacking GDNF.}
\newblock Nature 382: 73--76.
\input{Pichel:1996en}

\bibitem{Pepicelli:1997jz}
Pepicelli CV, Kispert A, Rowitch DH, McMahon AP (1997) {GDNF induces branching
  and increased cell proliferation in the ureter of the mouse.}
\newblock Developmental Biology 192: 193--198.
\input{Pepicelli:1997jz}

\bibitem{Sanchez:1996cy}
S{\'a}nchez MP, Silos-Santiago I, Fris{\'e}n J, He B, Lira SA, et~al. (1996)
  {Renal agenesis and the absence of enteric neurons in mice lacking GDNF.}
\newblock Nature 382: 70--73.
\input{Sanchez:1996cy}

\bibitem{Tang:2002it}
Tang MJ, Cai Y, Tsai SJ, Wang YK, Dressler GR (2002) {Ureteric bud outgrowth in
  response to RET activation is mediated by phosphatidylinositol 3-kinase.}
\newblock Developmental Biology 243: 128--136.
\input{Tang:2002it}

\bibitem{Tang:1998vf}
Tang MJ, Worley D, Sanicola M, Dressler GR (1998) {The RET-glial cell-derived
  neurotrophic factor (GDNF) pathway stimulates migration and chemoattraction
  of epithelial cells.}
\newblock J Cell Biol 142: 1337--1345.
\input{Tang:1998vf}

\bibitem{Sariola:2003jo}
Sariola H, Saarma M (2003) {Novel functions and signalling pathways for GDNF.}
\newblock Journal of cell science 116: 3855--3862.
\input{Sariola:2003jo}

\bibitem{Schuchardt:1994hg}
Schuchardt A, D'Agati V, Larsson-Blomberg L, Costantini F, Pachnis V (1994)
  {Defects in the kidney and enteric nervous system of mice lacking the
  tyrosine kinase receptor Ret.}
\newblock Nature 367: 380--383.
\input{Schuchardt:1994hg}

\bibitem{Menshykau:nMxfL07C}
Menshykau D, Iber D (2013) {Kidney branching morphogenesis under the control of a
  ligand-receptor-based Turing mechanism}.
\newblock Physical Biology 10: 046003
\input{Menshykau:nMxfL07C}

\bibitem{Menshykau:2012kg}
Menshykau D, Kraemer C, Iber D (2012) {Branch Mode Selection during Early Lung
  Development.}
\newblock Plos Computational Biology 8: e1002377.
\input{Menshykau:2012kg}

\bibitem{Schnakenberg:1979td}
Schnakenberg J (1979) {Simple chemical reaction systems with limit cycle
  behaviour.}
\newblock Journal of theoretical biology 81: 389--400.
\input{Schnakenberg:1979td}

\bibitem{Gierer:1972vq}
Gierer A, Meinhardt H (1972) {A theory of biological pattern formation.}
\newblock Kybernetik 12: 30--39.
\input{Gierer:1972vq}

\bibitem{Turing:1952p868}
Turing A (1952) {The chemical basis of morphogenesis}.
\newblock Phil Trans Roy Soc Lond B237: 37--72.
\input{Turing:1952p868}

\bibitem{Gleghorn:2012el}
Gleghorn JP, Kwak J, Pavlovich AL, Nelson CM (2012) {Inhibitory morphogens and
  monopodial branching of the embryonic chicken lung.}
\newblock Developmental dynamics : an official publication of the American
  Association of Anatomists .
\input{Gleghorn:2012el}

\bibitem{Nelson:2006gn}
Nelson CM, Vanduijn MM, Inman JL, Fletcher DA, Bissell MJ (2006) {Tissue
  geometry determines sites of mammary branching morphogenesis in organotypic
  cultures.}
\newblock Science 314: 298--300.
\input{Nelson:2006gn}

\bibitem{Riccio:2012ew}
Riccio PN, Michos O (2012) {Dissecting and culturing and imaging the mouse
  urogenital system.}
\newblock Methods in molecular biology (Clifton, NJ) 886: 3--11.
\input{Riccio:2012ew}

\bibitem{D'Errico:2012:Online}
D'Errico R (2012).
\newblock Interparc function.
\newblock
  \urlprefix\url{http://www.mathworks.in/matlabcentral/fileexchange/34874-inte%
rparc}.
\input{D'Errico:2012:Online}

\bibitem{D'Errico:2012dist}
D'Errico R (2012).
\newblock Normal distance function.
\newblock
  \urlprefix\url{http://www.mathworks.com/matlabcentral/fileexchange/34869-dis%
tance2curve}.
\input{D'Errico:2012dist}

\bibitem{cutress2010}
Cutress IJ, Dickinson EJF, Compton RG (2010) {Analysis of commercial general
  engineering finite element software in electrochemical simulations.}
\newblock J Electroanal Chem 638: 76--83.
\input{cutress2010}

\bibitem{Carin2006}
Carin M (2006) {Numerical Simulation of Moving Boundary Problems with the ALE
  Method: Validation in the Case of a Free Surface and a Moving Solidification
  Front.}
\newblock Excert from the Proceedings of the COMSOL Conference .
\input{Carin2006}

\bibitem{Thummler2007}
Thummler V, Weddemann A (2007) {Computation of Space-Time Patterns via ALE
  Methods.}
\newblock Excert from the Proceedings of the COMSOL Conference .
\input{Thummler2007}

\bibitem{Weddemann2008}
Weddemann A, Thummler V (2008) {Stability Analysis of ALE-Methods for
  Advection-Diffusion Problems.}
\newblock Excert from the Proceedings of the COMSOL Conference .
\input{Weddemann2008}

\bibitem{Menshykau:2012vg}
Menshykau D, Iber D (2012) {Simulating Organogenesis with Comsol: Interacting
  and Deforming Domains}.
\newblock Proceedings of COMSOL Conference 2012 .
\input{Menshykau:2012vg}

\bibitem{Germann:bT_kMV7D}
Germann P, Menshykau D, Tanaka S, Iber D (2011) {Simulating Organogensis in
  COMSOL}.
\newblock In: Proceedings of COMSOL Conference 2011.
\input{Germann:bT_kMV7D}

\bibitem{Pachnis:1993um}
Pachnis V, Mankoo B, Costantini F (1993) {Expression of the c-ret
  proto-oncogene during mouse embryogenesis.}
\newblock Development (Cambridge, England) 119: 1005--1017.
\input{Pachnis:1993um}

\bibitem{Bookstein:1989wl}
Bookstein FL (1989) {Principal warps: Thin-plate splines and the decomposition
  of deformations}.
\newblock Pattern Analysis and Machine Intelligence .
\input{Bookstein:1989wl}

\end{thebibliography}

\section*{Appendix}

S.A. is a Master Student. S.A. carried out the computational analysis under supervision of D.M. O.M. obtained the experimental data. D.I. conceived the project and wrote the paper together with D.M. and S.A.

\end{document}